# Longitudinal and transverse components of a vector field*

A. M. Stewart

Emeritus Faculty
The Australian National University,
Canberra, ACT 0200, Australia.

**Abstract**

A unified account, from a pedagogical perspective, is given of the longitudinal and transverse projective delta functions proposed by Belinfante and of their relation to the Helmholtz theorem for the decomposition of a three-vector field into its longitudinal and transverse components. It is argued that the results are applicable to fields that are time-dependent as well as fields that are time-independent.

## 1. INTRODUCTION

Belinfante [1] first discussed the longitudinal (l) and transverse (t) delta functions $\delta_l^{ij}(\mathbf{r} - \mathbf{r}')$ and $\delta_t^{ij}(\mathbf{r} - \mathbf{r}')$ which project out the longitudinal $\mathbf{A}_l(\mathbf{r})$ and transverse $\mathbf{A}_t(\mathbf{r})$ components of a three-vector field $\mathbf{A}(\mathbf{r}) = \mathbf{A}_l(\mathbf{r}) + \mathbf{A}_t(\mathbf{r})$ defined by the relations

$$\nabla \times \mathbf{A}_l(\mathbf{r}) = \mathbf{0} \quad \text{and} \quad \nabla \cdot \mathbf{A}_t(\mathbf{r}) = 0 \quad , \quad (1)$$

$\nabla$ being the gradient operator with respect to $\mathbf{r}$ and a symbol in bold font denotes a three-vector:

$$A_l^i(\mathbf{r}) = \sum_j \int dV' \, \delta_l^{ij}(\mathbf{r} - \mathbf{r}') A^j(\mathbf{r}') \quad (2)$$

$$A_t^i(\mathbf{r}) = \sum_j \int dV' \, \delta_t^{ij}(\mathbf{r} - \mathbf{r}') A^j(\mathbf{r}') \quad , \quad (3)$$

where $dV'$ is the volume element at $\mathbf{r}'$.

      These projective delta functions are used in electrodynamics [1, 27] and particularly in the quantisation of the electromagnetic field in the Coulomb gauge [2-4], a gauge that is extensively used in quantum chemistry and condensed matter physics [3]. They are closely related to Helmholtz's theorem, which also provides expressions for the longitudinal and transverse components of a vector field [5-8, 10, 14, 28]. Although the projective delta functions and Helmholtz's theorem are so closely related, they are not treated together in most texts. It seems worthwhile to give a systematic, straightforward and unified derivation of these two results on the basis of elementary differential vector relations [9] without the introduction of more advanced notions such as the inverse of the Laplacian operator [6].

In section 2, a derivation is given of the longitudinal and transverse components of a vector field and in section 3, the delta function projection operators of Belinfante [1] are obtained from the results of section 2 and their properties surveyed. In section 4 the longitudinal and transverse components of a vector field are transformed into their Helmholtz form and in sections 5 and 6, the results are applied to some simple problems in electrodynamics. In section 7, arguments are given that the Helmholtz theorem and the projective delta functions apply to fields that depend on time as well as to those that are time-independent. The appendix gives a derivation and extension of the Frahm [11] relation Eq. (17).

## 2. DERIVATION OF THE LONGITUDINAL AND TRANSVERSE COMPONENTS

Starting from the tautology

$$\mathbf{A}(\mathbf{r}) = \int dV' \, \delta(\mathbf{r} - \mathbf{r}') \mathbf{A}(\mathbf{r}') \tag{4}$$

where $\delta(\mathbf{r} - \mathbf{r}')$ is the three dimensional Dirac delta function and $\mathbf{A}(\mathbf{r})$ is any three-vector field, and using the standard identity [5]

$$\delta(\mathbf{r} - \mathbf{r}') = -\nabla^2 (1/4\pi |\mathbf{r} - \mathbf{r}'|) \tag{5}$$

we have

$$\mathbf{A}(\mathbf{r}) = -\int dV' \, \mathbf{A}(\mathbf{r}') \nabla^2 (1/4\pi |\mathbf{r} - \mathbf{r}'|) \quad . \tag{6}$$

If we use the relation $\nabla^2 [\mathbf{A}(\mathbf{r}') f(\mathbf{r})] = \mathbf{A}(\mathbf{r}') \nabla^2 f(\mathbf{r})$ we get

$$\mathbf{A}(\mathbf{r}) = -\int dV' \, \nabla^2 [\mathbf{A}(\mathbf{r}')/4\pi |\mathbf{r} - \mathbf{r}'|] \quad . \tag{7}$$

Next, using the identity for any three-vector $\mathbf{E}(\mathbf{r})$,

$$\nabla^2 \mathbf{E}(\mathbf{r}) = -\nabla \times \nabla \times \mathbf{E} + \nabla(\nabla \cdot \mathbf{E}) \tag{8}$$

we obtain from each term of Eq. (8)

$$\mathbf{A}(\mathbf{r}) = \mathbf{A}_l(\mathbf{r}) + \mathbf{A}_t(\mathbf{r}) \tag{9}$$

where

$$\mathbf{A}_l(\mathbf{r}) = -\nabla \int dV' \, \nabla \cdot [\mathbf{A}(\mathbf{r}')/4\pi |\mathbf{r} - \mathbf{r}'|] \tag{10}$$

and

$$\mathbf{A}_t(\mathbf{r}) = \nabla \times \int dV' \, \nabla \times [\mathbf{A}(\mathbf{r}')/4\pi |\mathbf{r} - \mathbf{r}'|] \quad . \tag{11}$$

It can be seen that $\mathbf{A}_l(\mathbf{r})$ and $\mathbf{A}_t(\mathbf{r})$ are respectively the longitudinal and transverse components of the field $\mathbf{A}(\mathbf{r})$ since from standard vector relations it follows that they obey Eqs. (1).

## 3. DERIVATION OF THE LONGITUDINAL AND TRANSVERSE DELTA FUNCTIONS AND THEIR PROPERTIES

First we consider the longitudinal component. It follows from Eq. (10) that

$$\mathbf{A}_l(\mathbf{r}) = -\int dV' \, \nabla[\mathbf{A}(\mathbf{r}')\cdot\nabla(1/4\pi|\mathbf{r}-\mathbf{r}'|)] \quad . \tag{12}$$

Applying the vector identity $\nabla[\mathbf{A}(\mathbf{r}')\cdot\mathbf{D}(\mathbf{r})] = \mathbf{A}\times(\nabla\times\mathbf{D}) + (\mathbf{A}\cdot\nabla)\mathbf{D}$ to Eq. (12) with $\mathbf{D}(\mathbf{r}) = \nabla(1/4\pi|\mathbf{r}-\mathbf{r}'|)$, and noting that because $\mathbf{D}$ is a gradient the first term vanishes, Eq. (12) becomes

$$\mathbf{A}_l(\mathbf{r}) = -\int dV' \, [\mathbf{A}(\mathbf{r}')\cdot\nabla]\nabla(1/4\pi|\mathbf{r}-\mathbf{r}'|) \quad . \tag{13}$$

Expressing this in vector components

$$A_l^i(\mathbf{r}) = \sum_j \int dV' \, [A^j(\mathbf{r}')\partial/\partial x^j](\partial/\partial x^i)(-1/4\pi|\mathbf{r}-\mathbf{r}'|) \tag{14}$$

and comparing with Eq. (2) we find

$$\delta_l^{ij}(\mathbf{r}-\mathbf{r}') = -\frac{\partial}{\partial x^i}\frac{\partial}{\partial x^j}\frac{1}{4\pi|\mathbf{r}-\mathbf{r}'|} \quad . \tag{15}$$

This expression diverges as $\mathbf{r} \to \mathbf{r}'$ and has to be regularised as the limit as $\varepsilon \to 0$ of

$$\delta_l^{ij}(\mathbf{r}-\mathbf{r}') = -\frac{1}{4\pi}\frac{\partial}{\partial x^i}\frac{\partial}{\partial x^j}\frac{1}{\sqrt{(\mathbf{r}-\mathbf{r}')^2+\varepsilon^2}} \quad . \tag{16}$$

Differentiating this regularised expression, as carefully shown by [11, 12], we get

$$\delta_l^{ij}(\mathbf{r}-\mathbf{r}') = \frac{1}{3}\delta_{ij}\delta(\mathbf{r}-\mathbf{r}') + \frac{1}{4\pi|\mathbf{r}-\mathbf{r}'|^3}[\delta_{ij} - \frac{3(x^i-x'^i)(x^j-x'^j)}{|\mathbf{r}-\mathbf{r}'|^2}] \quad , \tag{17}$$

where $\delta_{ij}$ is the Kronecker delta. See the Appendix for a derivation of Eq. (17) and its generalisation. Eq. (5) is recovered by equating the traces of the right hand sides of (16) and (17).

Next, the transverse component Eq. (11) becomes, by the vector identity $\nabla\times[\mathbf{A}g] = (\nabla g)\times\mathbf{A} + g\nabla\times\mathbf{A}$ where $g$ is a scalar field,

$$\mathbf{A}_t(\mathbf{r}) = -\int dV' \, \nabla\times[\mathbf{A}(\mathbf{r}')\times\nabla(1/4\pi|\mathbf{r}-\mathbf{r}'|)] \quad . \tag{18}$$

With the use of another vector relation $\nabla \times [\mathbf{A}(\mathbf{r}') \times \mathbf{D}(\mathbf{r})] = \mathbf{A}(\nabla \cdot \mathbf{D}) - (\mathbf{A} \cdot \nabla)\mathbf{D}$, with the same $\mathbf{D}(\mathbf{r})$ as below Eq. (12), the first term of the right hand side of Eq. (18), when the identity is substituted into it, becomes the integral over $V'$ of $-\mathbf{A}(\mathbf{r}')\delta(\mathbf{r} - \mathbf{r}')$. The second term becomes identical to the right hand side of Eq. (13) giving, when compared to Eqs. (2, 3),

$$\delta_t^{ij}(\mathbf{r} - \mathbf{r}') = \delta_{ij}\delta(\mathbf{r} - \mathbf{r}') - \delta_l^{ij}(\mathbf{r} - \mathbf{r}') \qquad . \qquad (19)$$

It should be noted that Eq. (19) is derived, and not assumed on the basis of Eq. (9).

Using Eq. (17) the transverse projection operator may be expressed explicitly as

$$\delta_t^{ij}(\mathbf{r} - \mathbf{r}') = \frac{2}{3}\delta_{ij}\delta(\mathbf{r} - \mathbf{r}') - \frac{1}{4\pi|\mathbf{r}-\mathbf{r}'|^3}[\delta_{ij} - \frac{3(x^i - x'^i)(x^j - x'^j)}{|\mathbf{r}-\mathbf{r}'|^2}] \qquad . \qquad (20)$$

A number of properties follow from Eqs. (15) and (19) where the subscript *s* indicates either projection operator

$$\delta_s^{ij}(\mathbf{r} - \mathbf{r}') = \delta_s^{ji}(\mathbf{r} - \mathbf{r}') \quad , \qquad \delta_s^{ij}(\mathbf{r} - \mathbf{r}') = \delta_s^{ij}(\mathbf{r}' - \mathbf{r}) \qquad , \qquad (21)$$

$$\frac{\partial}{\partial x^j}\delta_l^{kl}(\mathbf{r} - \mathbf{r}') = \frac{\partial}{\partial x^k}\delta_l^{jl}(\mathbf{r} - \mathbf{r}') \qquad , \qquad (22)$$

confirming that $\nabla \times \mathbf{A}_l(\mathbf{r}) = \mathbf{0}$ and

$$\sum_i \frac{\partial}{\partial x^i}\delta_l^{ij}(\mathbf{r} - \mathbf{r}') = \frac{\partial}{\partial x^j}\delta(\mathbf{r} - \mathbf{r}') \quad , \qquad \sum_i \frac{\partial}{\partial x^i}\delta_t^{ij}(\mathbf{r} - \mathbf{r}') = 0 \qquad , \qquad (23)$$

confirming that $\nabla \cdot \mathbf{A}_t(\mathbf{r}) = 0$. By means of several partial integrations in Cartesian coordinates where the surface terms vanish because of the $1/|\mathbf{r}|$ factor it can be shown that both the operators in the forms (16) and (19) are idempotent

$$\delta_s^{ij}(\mathbf{r} - \mathbf{r}') = \sum_k \int dV'' \delta_s^{ik}(\mathbf{r} - \mathbf{r}'')\delta_s^{kj}(\mathbf{r}'' - \mathbf{r}') \qquad . \qquad (24)$$

By expressing $\mathbf{A}_l$ as a gradient $\nabla a$ and using the vector identity $\nabla \cdot (a\mathbf{C}_t) = \nabla a \cdot \mathbf{C}_t$ it can also be shown that

$$\int dV\, \mathbf{A}_l(\mathbf{r}) \cdot \mathbf{C}_t(\mathbf{r}) = 0 \qquad (25)$$

for any vectors $\mathbf{A}(\mathbf{r})$ and $\mathbf{C}(\mathbf{r})$ that vanish at infinity. By doing partial integrations in Cartesian coordinates over $x^i$ and $x^j$ in (15) and (19) and noting that the Fourier transform of $1/4\pi|\mathbf{r}|$ is $1/k^2$ it is found that the Fourier transforms of the projection operators

$$\delta_s^{ij}(\mathbf{k}) = \int dV\, e^{-i\mathbf{k}\cdot\mathbf{r}} \delta_s^{ij}(\mathbf{r}) \qquad (26)$$

are

$$\delta_l^{ij}(\mathbf{k}) = k^i k^j / \mathbf{k}^2 \qquad \text{and} \qquad \delta_t^{ij}(\mathbf{k}) = \delta_{ij} - k^i k^j / \mathbf{k}^2 \qquad . \qquad (27)$$

If the denominator of the second term of the right hand side of (17) is regularized in the same way as in (16) its integral over all space comes to $-\delta_{ij}/3$. Consequently

$$\int dV\, \delta_t^{ij}(\mathbf{r}) = \delta_{ij} \qquad \text{and} \qquad \int dV\, \delta_l^{ij}(\mathbf{r}) = 0 \qquad .$$

## 4. HELMHOLTZ EXPRESSIONS FOR THE LONGITUDINAL AND TRANSVERSE COMPONENTS

The longitudinal (12) and transverse (18) components may be transformed into the Helmholtz form by further manipulation. From Eq. (12), exchanging the primed and unprimed variables in the term following the second $\nabla$, we find

$$\mathbf{A}_l(\mathbf{r}) = \nabla \int dV'\, \mathbf{A}(\mathbf{r'}) \cdot \nabla'(1/4\pi |\mathbf{r} - \mathbf{r'}|) \qquad . \qquad (28)$$

Then, by using the identity

$$\nabla' \cdot [\mathbf{A}(\mathbf{r'})/|\mathbf{r} - \mathbf{r'}|] = \mathbf{A}(\mathbf{r'}) \cdot \nabla'(1/|\mathbf{r} - \mathbf{r'}|) + [\nabla' \cdot \mathbf{A}(\mathbf{r'})]/|\mathbf{r} - \mathbf{r'}| \qquad , \qquad (29)$$

we get

$$\mathbf{A}_l(\mathbf{r}) = \frac{\nabla}{4\pi} \int dV' \{\nabla' \cdot [\mathbf{A}(\mathbf{r'})/|\mathbf{r} - \mathbf{r'}|] - [\nabla' \cdot \mathbf{A}(\mathbf{r'})]/|\mathbf{r} - \mathbf{r'}|\} \qquad . \qquad (30)$$

By Gauss's theorem the first term of Eq. (30) becomes a surface integral

$$\nabla \int d\mathbf{S'} \cdot [\mathbf{A}(\mathbf{r'})(1/4\pi |\mathbf{r} - \mathbf{r'}|)] \qquad (31)$$

where $d\mathbf{S'}$ is the directed surface element $d\mathbf{S'} = \hat{\mathbf{R}}' R'^2 d\Omega'$ and $d\Omega'$ is the infinitesimal solid angle at radius $R'$. It can be seen that the surface integral will vanish if $\mathbf{A}$ itself vanishes at spatial infinity. In this case the longitudinal component takes the Helmholtz value

$$\mathbf{A}_l(\mathbf{r}) = -\nabla \int dV' [\nabla' \cdot \mathbf{A}(\mathbf{r'})](1/4\pi |\mathbf{r} - \mathbf{r'}|) \qquad . \qquad (32)$$

By exchanging $\nabla$ for $\nabla'$ the transverse component, Eq. (18), becomes

$$\mathbf{A}_t(\mathbf{r}) = \nabla \times \int dV'\, \mathbf{A}(\mathbf{r'}) \times \nabla'(1/4\pi |\mathbf{r} - \mathbf{r'}|) \qquad . \qquad (33)$$

Then, with the identity

$$\nabla' \times [\frac{\mathbf{A}(\mathbf{r'})}{|\mathbf{r} - \mathbf{r'}|}] = \frac{1}{|\mathbf{r} - \mathbf{r'}|} \nabla' \times \mathbf{A}(\mathbf{r'}) - \mathbf{A}(\mathbf{r'}) \times \nabla'(\frac{1}{|\mathbf{r} - \mathbf{r'}|}) \qquad , \qquad (34)$$

we get

$$\mathbf{A}_t(\mathbf{r}) = \frac{\nabla}{4\pi} \times \int dV' \left\{ \frac{[\nabla' \times \mathbf{A}(\mathbf{r}')]}{|\mathbf{r} - \mathbf{r}'|} - \nabla' \times [\frac{\mathbf{A}(\mathbf{r}')}{|\mathbf{r} - \mathbf{r}'|}] \right\} \qquad . \qquad (35)$$

The volume integral in the second term may be expressed as a surface integral

$$\int dV' \nabla' \times [\frac{\mathbf{A}(\mathbf{r}')}{|\mathbf{r} - \mathbf{r}'|}] = \int d\mathbf{S}' \times [\frac{\mathbf{A}(\mathbf{r}')}{|\mathbf{r} - \mathbf{r}'|}] \qquad . \qquad (36)$$

Noting the presence of the gradient that acts on the integral, it can be seen that the surface term vanishes if $\mathbf{A}(\mathbf{r})$ vanishes sufficiently fast at spatial infinity so the transverse field also attains its Helmholtz form

$$\mathbf{A}_t(\mathbf{r}) = \nabla \times \int dV' [\nabla' \times \mathbf{A}(\mathbf{r}')](1/4\pi |\mathbf{r} - \mathbf{r}'|) \qquad . \qquad (37)$$

The longitudinal and transverse components of the field $\mathbf{A}(\mathbf{r})$ are given by either Eqs. (10) and (11) or the Helmholtz forms Eqs. (32) and (37). They have been shown to be valid for the long-range radiation fields of electromagnetism [28].

## 5. ELECTROMAGNETIC VECTOR POTENTIAL IN THE COULOMB GAUGE

If $\mathbf{A}$ is the vector potential of the electromagnetic field then putting $\mathbf{B}(\mathbf{r}') = \nabla' \times \mathbf{A}(\mathbf{r}')$ into Eq. (37) we get

$$\mathbf{A}_t(\mathbf{r}) = \nabla \times \frac{1}{4\pi} \int dV' \frac{\mathbf{B}(\mathbf{r}')}{|\mathbf{r} - \mathbf{r}'|} \qquad (38)$$

or
$$\mathbf{A}_t(\mathbf{r}) = \frac{1}{4\pi} \int dV' \frac{\mathbf{B}(\mathbf{r}') \times (\mathbf{r} - \mathbf{r}')}{|\mathbf{r} - \mathbf{r}'|^3} \qquad . \qquad (39)$$

Eqs. (38-39) give the transverse part of the vector potential that is responsible for the $\mathbf{B}(\mathbf{r})$ field; it is manifestly gauge invariant and can be in principle computed from a given $\mathbf{B}$ field.

## 6. VECTOR POTENTIAL OF A STRAIGHT DIRAC STRING

As an illustration of the use of Eq. (39) we calculate the vector potential in the Coulomb gauge for the field $\mathbf{B}(\mathbf{r}) = \hat{\mathbf{z}}\Phi\delta(x)\delta(y)$ in a long thin solenoid (or a straight Dirac string) along the z-axis where $\Phi$ is the magnetic flux in the solenoid.

$$\mathbf{A}_t(\mathbf{r}) = \frac{\Phi}{4\pi} \int dx' dy' dz' \frac{\delta(x')\delta(y')\hat{\mathbf{z}} \times (\mathbf{r} - \mathbf{r}')}{|\mathbf{r} - \mathbf{r}'|^3} \qquad . \qquad (40)$$

Noting that $\mathbf{r}' = \hat{\mathbf{z}} z'$ we get

$$\mathbf{A}_t(\mathbf{r}) = \frac{\Phi}{4\pi} \int_{-\infty}^{\infty} dz' \frac{(\hat{\mathbf{y}} x - \hat{\mathbf{x}} y)}{|\mathbf{r} - \mathbf{r}'|^3} \qquad . \qquad (41)$$

The numerator of the integrand is a vector in the $\boldsymbol{\theta}$ direction of magnitude $(x^2 + y^2)^{1/2}$, the denominator is $[x^2 + y^2 + (z - z')^2]^{3/2}$. The integral then comes to

$$\mathbf{A}_t(\mathbf{r}) = \frac{\hat{\boldsymbol{\theta}} \Phi (x^2 + y^2)^{1/2}}{4\pi} \int_{-\infty}^{\infty} \frac{dz'}{[x^2 + y^2 + (z - z')^2]^{3/2}} \quad , \tag{42}$$

and, carrying out the integration, we get finally $\mathbf{A}_t(\mathbf{r}) = \hat{\boldsymbol{\theta}} \Phi / (2\pi |\mathbf{r}|)$, the result usually [7] obtained in a simpler way by equating the line integral of the vector potential to the flux enclosed in a circular path around the solenoid.

## 7. TIME-DEPENDENT FIELDS

The discussions in this paper have been conducted in terms of fields that are independent of time. A question that naturally arises is whether the Helmholtz theorem Eqs. (32) and (37) and the projective delta function relations Eqs. (17) and (20) are valid for vector fields that do depend on time $t$; in other words, whether $\mathbf{A}(\mathbf{r})$ in Eqs. (32) and (33) can be replaced by $\mathbf{A}(\mathbf{r}, t)$? That it can be has been assumed implicitly in applications of the Helmholtz theorem to the Coulomb gauge [16, 17], to the momentum and angular momentum of the electromagnetic field [19-22], to some quadratic identities in electromagnetism [23] and to quantum electrodynamic calculations in the Coulomb gauge [2, 3].

      This assumption is justified by the fact that the arguments used in this paper use relationships like Eqs. (5) and (8) that involve spatial coordinates only. The fields may be functions of other variables such as time but, as long as these other variables are held constant, the derivation is unchanged. Rohrlich [15] and Woodside [14, 25-25] have also arrived at this conclusion using related arguments.

      The issue appears in its most salient form when Eq. (39) is written with the time explicit:

$$\mathbf{A}_t(\mathbf{r}, t) = \frac{1}{4\pi} \int dV' \frac{\mathbf{B}(\mathbf{r}', t) \times (\mathbf{r} - \mathbf{r}')}{|\mathbf{r} - \mathbf{r}'|^3} \tag{43}$$

and it is seen that the transverse vector potential at time $t$ is given in terms of the magnetic field at the same time but at different positions. The expression is formally instantaneous but, because the vector potential is not a physically measurable quantity, this is not a matter for concern [15-18]. It is readily found [17] that by taking the curl of Eq. (43) the magnetic field $\mathbf{B}(\mathbf{r}, t)$ is obtained.

## 8. SUMMARY

A unified derivation, from a pedagogical perspective, has been given of the Helmholtz theorem for the decomposition of vector fields and of the projective delta functions of Belinfante [1] on the basis of the relation describing the partial derivatives of the Coulomb potential. The results are shown to be valid for fields that depend on time. Some applications to electromagnetism are discussed briefly.

**APPENDIX**

In this appendix we derive (17) [11] and obtain some generalisations of it. It follows from simple differentiation that

$$\frac{\partial}{\partial x^i}\frac{\partial}{\partial x^j}\frac{1}{\sqrt{r^2+\varepsilon^2}} = \frac{3x^i x^j - \delta_{ij}r^2}{(r^2+\varepsilon^2)^{5/2}} - \frac{\delta_{ij}\varepsilon^2}{(r^2+\varepsilon^2)^{5/2}} \tag{A1}$$

where $r^2 = x^2 + y^2 + z^2$ and $x^i$ is one of $\{x, y, z\}$. Integrate the second term on the right over $d^3r$ to give, using the substitution $t = r/\varepsilon$,

$$\int_0^\infty dr \frac{4\pi r^2 \varepsilon^2}{(r^2+\varepsilon^2)^{5/2}} = \int_0^\infty dt \frac{4\pi t^2}{(1+t^2)^{5/2}} = \frac{4\pi}{3} \tag{A2}$$

The integral is independent of $\varepsilon$, provided that $\varepsilon$ is finite and, as $\varepsilon$ becomes smaller, the second term of (A1) becomes more peaked at $\mathbf{r} = 0$ so, as $\varepsilon \to 0$, the second term approaches a delta function and (A1) may be written

$$\frac{\partial}{\partial x^i}\frac{\partial}{\partial x^j}\frac{1}{r} = -\frac{4\pi}{3}\delta_{ij}\delta(\mathbf{r}) - \frac{1}{r^3}(\delta_{ij} - \frac{3x^i x^j}{r^2}) \tag{A3}$$

The appropriate limiting procedure in integrating any function in product with (A3) is to do the angular integrations before the radial integration [12].

When $n$ is an integer equal to or greater than zero, two generalisations of (A3) may be obtained from the differentiation

$$\frac{\partial}{\partial x^i}\frac{\partial}{\partial x^j}\frac{1}{r^n} = \frac{n(n-1)x^i x^j}{r^{n+4}} + \frac{n}{r^{n-1}}\left(\frac{\partial}{\partial x^i}\frac{\partial}{\partial x^j}\frac{1}{r}\right) \tag{A4}$$

which gives

$$\frac{\partial}{\partial x^i}\frac{\partial}{\partial x^j}\frac{1}{r^n} = -n\left[\frac{4\pi}{3}\frac{\delta_{ij}\delta(\mathbf{r})}{r^{n-1}} + \frac{\delta_{ij} - (n+2)x^i x^j / r^2}{r^{n+2}}\right], \tag{A5}$$

and

$$\frac{\partial}{\partial x^i}\frac{\partial}{\partial x^j}\frac{x^g}{r^n} = \delta_{gj}\frac{\partial}{\partial x^i}\frac{1}{r^n} + \delta_{gi}\frac{\partial}{\partial x^j}\frac{1}{r^n} + x^g\frac{\partial}{\partial x^i}\frac{\partial}{\partial x^j}\frac{1}{r^n} \tag{A6}$$

or

$$\frac{\partial}{\partial x^i}\frac{\partial}{\partial x^j}\frac{x^g}{r^n} = -\delta_{gj}n\frac{x^i}{r^{n+2}} - \delta_{gi}n\frac{x^j}{r^{n+2}} + x^g\frac{\partial}{\partial x^i}\frac{\partial}{\partial x^j}\frac{1}{r^n} \tag{A7}$$

This leads to

$$\frac{\partial}{\partial x^i}\frac{\partial}{\partial x^j}\frac{x^g}{r^n} = -nx^g\left[\frac{4\pi}{3}\frac{\delta_{ij}\delta(\mathbf{r})}{r^{n-1}} + \frac{\delta_{ij}-(n+2)x^i x^j/r^2}{r^{n+2}}\right] - n\frac{\delta_{gi}x^j + \delta_{gj}x^i}{r^{n+2}} \tag{A8}$$